\documentclass[usegraphicx,usenatbib,onecolumn]{mn2e}
\usepackage{bm}

\begin{document}
\title{S-Z power spectrum produced by 
primordial magnetic fields}

\author[Tashiro, H. et al.]
{Hiroyuki Tashiro$^1$$^2$, and Naoshi Sugiyama$^3$$^4$\\
$^1$Institut d'Astrophysique Spatiale (IAS), B\^atiment 121, Universit\'e Paris-Sud 11, 
Orsay, F-91405, France\\
$^2$Center for Particle Physics and Phenomenology (CP3), 
Universite catholique de Louvain,\\
Chemin du Cyclotron, 2, B-1348 Louvain-la-Neuve, Belgium\\
$^3$Department of Physics and Astrophysics, Nagoya University, 
  Chikusa, Nagoya 464-8602, Japan\\
$^4$Institute for Physics and Mathematics of the Universe, 
University of Tokyo,\\ 
 5-1-5 Kashiwa-no-Ha, Kashiwa,
Chiba, 277-8582, Japan}

\date{\today}

\maketitle

\begin{abstract}

Primordial magnetic fields generated in the very early universe are
one of the candidates for the origin of magnetic fields observed in
galaxy clusters. After recombination, the Lorentz force acts on the
residual ions and electrons to generate density fluctuations of
baryons.  Accordingly these fluctuations induce the early formation of
dark halos which cause the Sunyaev-Zel'dovich (S-Z) effect in cosmic
microwave background radiation.  This additional S-Z effect due to
primordial magnetic fields amplifies the angular power spectrum of
cosmic microwave temperature anisotropies on small scales.  This
amplification depends on the comoving amplitude and the power law
index of the primordial magnetic fields spectrum.  Comparing with the
small scale CMB observations, we obtained the constraints on the
primordial magnetic fields,  
i.e., $B \la 2.0 $ nGauss for $n=-2.9$ or $B
\la 1.0 $ nGauss for $n=-2.6$ , where $B $ is the comoving amplitude of
magnetic fields at $h^{-1}$ Mpc and $n$ is the power law index.  
Future S-Z measurements have the potential to give  
constraints tighter than those from temperature anisotropies
and polarization of cosmic microwave background induced by the
magnetic fields at the recombination epoch.
\end{abstract}
\maketitle

\section{introduction}

Many observations indicate the existence of large-scale magnetic fields 
associated with galaxies and galaxy clusters.
These magnetic fields typically have strengths of a few $\mu $Gauss 
and large coherence lengths, i.e., a few kpc for galaxies and a few tens of kpc 
for galaxy clusters~\citep{magobs}.
However, the origin of such magnetic fields is not understood clearly, while
many generation processes have been proposed.   

The generally accepted idea is an astrophysical dynamo scenario.
Very tiny seed magnetic fields are generated 
in stars and supernova explosions by astrophysical processes such as Biermann
battery, and the produced seed magnetic fields 
are amplified by the dynamo process in astrophysical objects.
Finally these magnetic fields are spread into the inter-galactic medium by 
supernova winds or active galactic nuclei jets
~\citep{widrow,brandenburg-subramanian}.
However, there are two major problems remaining in this scenario.
The first problem is the efficiency of the dynamo process in the expanding 
universe.
Recent observations
suggest the existence of $\mu$Gauss magnetic fields in high redshift
galaxies~\citep{maghighz}.  These galaxies may be dynamically too
young to explain the existence of such magnetic fields by the dynamo process.
The second problem concerns large coherence lengths.
It is particularly difficult to explain observed magnetic fields with very
large coherent scales in galaxy clusters~\citep{clustermag1,clustermag2}. 

Aside from this astrophysical scenario, there are alternative
scenarios in which magnetic fields are generated in the early
universe, e.g., inflation epoch or cosmological phase transitions such
as QCD or electroweak. In these scenarios, there is the potential to
obtain nano Gauss primordial magnetic fields.  Such strength is
sufficient to explain $\mu$Gauss magnetic fields observed at
present without the dynamo process because the adiabatic compression
due to the structure formation can easily amplify primordial magnetic
fields by a factor of $\sim 10^3$.  However, if the seed magnetic fields
generated in the early universe are too weak, the dynamo process is
required even in these scenarios while the coherence length could be
very large, unlike the astrophysical processes.  For a detailed review,
see~\citet{giova}.

If primordial magnetic fields existed in the early universe, these
fields left traces of their existences in various cosmological
phenomena, e.g., big bang nucleosynthesis (BBN), temperature
anisotropies and polarization of cosmic microwave background (CMB), or
large scale structure formations.  From these traces, we can set
observational constraints on primordial magnetic fields.  These
constraints give us clues to the origin of large scale magnetic
fields, as well as when and how primordial magnetic fields were generated,
because the strength and the coherence length of primordial magnetic
fields depend on the generation process.

Let us first summarize BBN constraint.  
Since the primordial magnetic fields enhanced the cosmological expansion rate
through the contribution of the energy density of primordial magnetic fields 
to the total energy density of the universe, 
the existence of primordial magnetic fields with sufficient strength may modify 
the abundance of light elements. 
The constraint on the magnetic field strength from BBN
is $B_0 \la 7 \times 10^{-5}$Gauss 
where $B_0$ is the total comoving magnetic field 
strength~\citep{bbnconstraint1,bbnconstraint2}.

Primordial magnetic fields produce CMB temperature anisotropies.
Particularly, before recombination, primordial magnetic fields induce
the vorticity of a baryon fluid by the Lorentz force.  The induced
vorticity generates CMB temperature anisotropies through the Doppler
effect \citep{subramanian-barrow-1998}.  From the Wilkinson Microwave Anisotropy Probe 
(WMAP) data, the
constraint on the primordial magnetic fields with $1$Mpc--$100$Mpc is
$B_0 \la 10^{-8}$Gauss
\citep{mack-kahniashvili-02,lewis,tashiro-nonlinear,yamazaki-ichiki}.
Moreover, this vorticity generates CMB B-mode (parity odd) polarization
as well as E-mode (parity even)
polarization~\citep{subramanian-seshadri,tashiro-nonlinear}.  In
particular, B-modes are less contaminated by other sources than
E-modes so that we expect to obtain stringent limits on the
primordial magnetic fields by future observations of CMB B-modes.

After recombination, there are two main effects of primordial magnetic
fields on the universe.  One is the modification of the thermal
evolution of baryons~\citep{sethi-subramanian}.  Through the
dissipation of primordial magnetic fields, primordial magnetic fields
increase the baryon temperature after thermal decoupling of baryons
from CMB.  This dissipation is caused by the ambipolar diffusion and
the direct cascade decay of small scale magnetic fields.  The other
effect is the generation of density fluctuations
~\citep{wasserman,kim-olinto,gopal-sethi}.  The motion of ionized
baryons induced by magnetic fields produces additional density
fluctuations.  These fluctuations induce density fluctuations of
neutral baryons and dark matter through the gravitational force.  The
magnetic tension and pressure are more effective on small scales where
the entanglements of magnetic fields are larger.  Therefore, if
primordial magnetic fields existed, it is expected that there is
additional power in the density power spectrum, on small scales, which
induces the early structure formation.  These effects, modification of
baryon thermal history and generation of additional density
fluctuations, impact the reionization process.  Therefore, it
is possible to set constraints on primordial magnetic fields from the
measurement of the optical depth
\citep{sethi-subramanian,tashiro-early} and the observation of 21 cm
lines \citep{tashiro-21cm}.

In this paper, we investigate the effect of primordial magnetic fields
on the Sunyaev-Zel'dovich (S-Z) angular power spectrum.  The S-Z
effect occurs when CMB photons passing galaxy clusters are scattered
by hot electron gas in galaxy clusters \citep{szeffect-1972}.  Due
to the scattering, the CMB spectrum suffers distortion from the
blackbody shape.  The amount of distortion depends on the temperature
and the number density of hot electron gas.  In the low frequency
limit, i.e., the Rayleigh-Jeans part, this distortion causes
decrease in temperature, which is observed as the temperature
anisotropies in the CMB sky.  Since the distribution of hot electron
gas follows that of dark matter halos, the S-Z angular power
spectrum traces the dark matter halo distribution which could be
enhanced by primordial magnetic fields.  Moreover, it is known that
the S-Z effect is an ideal probe for the high redshift clusters/dark
halos because the strength of the S-Z signal does not depend on
redshift of the object, which is not the case for X-ray brightness
temperature or the gravitational lensing effect.  Since the
primordial magnetic fields induce structure formation in the early
epoch, we can conclude that the S-Z power spectrum can be used as a unique
probe for the primordial mangetic fields.


This paper is organized as follows.  In Sec. II, we discuss the
density fluctuations due to primordial magnetic fields. 
In Sec. III, we summarize the calculation of 
the angular power spectrum of the S-Z effect.
In Sec. IV, we show our results and discuss the constraint on primordial magnetic fields
from the S-Z power spectrum.
In Sec. V, we give the conclusion of this paper.
Throughout the paper, we take 3-yr WMAP results for the cosmological parameters, i.e.,
$h=0.70$ $(H_0=h \times 100 ~{\rm Km/s \cdot Mpc})$, $T_0 = 2.725$ K,
$\Omega _{\rm b} =0.044$, $\Omega_{\rm m}  =0.26$ \citep{wmap-spergel-07}
and we assume $\sigma_8 =0.8$.
We normalize the value of the velocity of light to 1.

\section{density fluctuations due to primordial magnetic fields}\label{density}

In this section, we calculate the density fluctuations produced by primordial magnetic fields. 
Let us make some assumptions about primordial magnetic fields at first.
Since the length scales which we are interested in are large,
the back-reaction of the fluid velocity is small.
Therefore, it is an assumption in this paper that
primordial magnetic fields are frozen in cosmic baryon fluids, 
\begin{equation}
{\bm B} (t,{\bm x}) ={ {\bm B} _0({\bm x}) \over a^2(t)},
\end{equation}
where $B_0({\bm x})$ is the comoving strength of primordial magnetic fields 
and $a(t)$ is the scale factor which is normalized as $a(t_0)=1$ at the present time, $t_0$. 
For simplicity, we assume that primordial magnetic fields are
statistically homogeneous and isotropic and have the power law spectrum with the 
power law index $n$,
\begin{equation}
\langle B_{0i}({\bm k_1} ) B_{0j} ^* ({\bm k_2} )\rangle= {(2 \pi)^3 \over 2}
\delta ({\bm k_1}-{\bm k_2}) \left(\delta_{ij}-{k_{1i} k_{2j} \over k_1^2 } \right) B^2 _{\rm n}
\left ({k \over k_{\rm n}} \right)^{n},
\label{eq:powerlaw}
\end{equation}
where $\langle ~ \rangle$ denotes the ensemble average,  
$B_{0i}({\bm k} )$ are Fourier components of $B_{0i}({\bm x})$, $k_{\rm n}$ 
is the wave number of an arbitrary normalized scale and 
$B_{\rm n}$ is the magnetic field strength at $k_{\rm n}$.

Our interest is to constrain the magnetic field strength on a certain scale in the real space.
Therefore, we have to convolve the power spectrum with 
a Gaussian filter transformation of a comoving radius $\lambda$,
in order to get the magnetic field strength in the real space,
\begin{equation}
B_\lambda ^2 \equiv \langle B_{0i}({\bm x} ) B_{0i}  ({\bm x} )\rangle |_\lambda =
{1 \over (2 \pi)^3} \int d^3 k B^2 _{\rm n}
\left ({k \over k_{\rm n}} \right)^{n} 
\left | \exp \left(- {\lambda ^2 k^2 \over 2} \right) \right | ^2.
\label{eq:mag-window}
\end{equation}
Substituting Eq.~(\ref{eq:powerlaw}) to Eq.~(\ref{eq:mag-window}),
we can associate $B_\lambda$ with $B_{\rm n}$,
\begin{equation}
B_\lambda ^2 ={B_{\rm n} ^2 \over (2 \pi)^2 \lambda^3 } ({ k_{\rm n} \lambda })^{-n}
\Gamma((n+3)/2).
\label{eq:normalizedmag}
\end{equation}
We take $h^{-1}~$Mpc as $\lambda$ throughout our paper.

Primordial magnetic fields produce vorticity in a cosmic fluid.
This vorticity is damped by the interaction between electrons and photons
around the recombination epoch.
This damping causes the dissipation of primordial magnetic fields
and causes a sharp cutoff on the power spectrum of primordial magnetic fields. 
The cutoff scale $1/k_{\rm c}$ after the recombination epoch is decided by
\citep{jedamzik-katalinic-98,subramanian-barrow-98},
\begin{equation}
k_{\rm c} ^{-2} = V_{\rm A}^2  \int ^{t_{\rm r}} {l_\gamma \over a^2 (t)} dt,
\label{eq:def-cut}
\end{equation}
where
$t_{\rm r}$ is the recombination time and $l_\gamma$ is the mean free path of photons,
which is described with the electron number density $n_e$ 
and the Thomson cross section $\sigma_T$ as $l_\gamma = 1/ n_e \sigma_T$.
In Eq.~(\ref{eq:def-cut}), $V_{\rm A}$ is the effective Alfv\'en velocity at the cutoff scale,
$V_{\rm A} = B_{\rm c} /\sqrt{4 \pi \rho_{\rm r}}$, where $\rho_{\rm r}$ is the radiation energy density
and 
$B_{\rm c}$ is the effective magnetic fields at the cutoff scale, which is obtained 
by smoothing primordial magnetic fields.
In the case of the power-law spectrum of primordial magnetic fields,
the $B_{\rm c}$ is given by \citep{mack-kahniashvili-02}
\begin{equation}
B_{\rm c} = B_\lambda \left ({ k_{\rm c} \over k_\lambda }\right) ^{(n+3)/2}.
\end{equation}
Assuming the matter dominated epoch,
we can obtain the relation between $k_{\rm c}$ and $B_{\lambda}$ as 
\begin{equation}
k_{\rm c} = \left[ 143 \left ( {B_{\lambda} \over 1 {\rm nG}} \right ) ^{-1}
\left ( { h  \over 0.7} \right )^{1/2} 
\left ({ h^2 \Omega_{\rm b} \over 0.021} \right )^{1/2} \right ]^{2/n+5} {\rm Mpc}^{-1} .
\label{eq:cutoff}
\end{equation}

\begin{figure}
  \begin{center}
    \includegraphics[keepaspectratio=true,height=50mm]{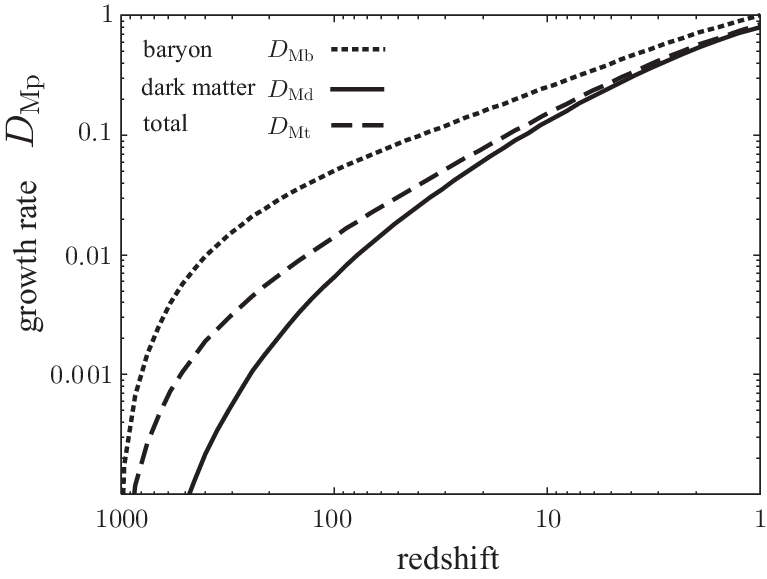}
  \end{center}
  \caption{Growth rates induced by primordial magnetic fields at given redshifts.
  The dotted and solid lines represent the growth rates for baryons and for 
  dark matter, respectively. We also plot the growth rate for total matter as the dashed line.
  All growth rates are normalized as $D_{{\rm M} \rm b}=1$ at $z=1$.}
    \label{fig:transfer}
\end{figure}

Primordial magnetic fields affect motions of ionized baryons by the Lorentz force
even after recombination \citep{wasserman}.
Although the residual ionized baryon rate to total baryons is small after recombination,
the interaction between ionized and neutral baryons is strong
in those redshifts that we are interested in.
Therefore, we can assume baryons as a MHD fluid.
Using the MHD approximation,
we can write the evolution equations of density fluctuations with primordial magnetic fields as, 
\begin{equation}
{\partial^2  \delta_{\rm b} \over \partial t^2} = -2 {\dot a \over a}
{\partial \delta_{\rm b} \over \partial t }
+4 \pi G (\rho _{\rm b} \delta_{\rm b} + \rho _{\rm d} \delta_{\rm d} ) 
+ S(t,{\bm x}),
\label{eq:baryon-den}
\end{equation}
\begin{equation}
S(t,{\bm x})={ \nabla \cdot \left( (\nabla \times {\bm B}_{0} ({\bm x})) 
\times  {\bm B}_{0} ({\bm x}) \right) 
\over 4 \pi \rho_{{\rm b} 0} a^3 (t) },
\end{equation}
\begin{equation}
{\partial^2  \delta_{\rm d} \over \partial t^2} = -2 {\dot a \over a}
{\partial \delta_{\rm d} \over \partial t }
+4 \pi G (\rho _{\rm b} \delta_{\rm b} + \rho _{\rm d} \delta_{\rm d} ),
\label{eq:dm-den}
\end{equation}
where 
$\rho_{\rm b}$ and $\rho_{\rm d}$ are the baryon density and the dark matter density, 
and $\delta_{\rm b}$ and $\delta_{\rm dm}$ are   
the density contrast of baryons and dark matter, respectively.
The source term in Eq.~(\ref{eq:dm-den}) 
is only the gravitational potential like that in the standard cosmology case, 
without primordial magnetic fields,
while other source term caused by magnetic fields is added in Eq.~(\ref{eq:baryon-den}).
The solutions of Eqs.~(\ref{eq:baryon-den}) and (\ref{eq:dm-den}) can be given by
\begin{equation}
\delta_{\rm p} = D_{{\rm S}\rm p}(t) \delta_{\rm p} (t_{\rm  i}) 
+D_{{\rm M} \rm p}(t)  t_{\rm  i} ^2 S( t_{\rm i}, {\bm x}),
\label{eq:density-solution}
\end{equation}
where $\rm p$ denotes $\rm b$ for baryons and $\rm d$ for dark matter.
Here $D_{{\rm S} \rm p}(t)$ corresponds to the growth rate of each component 
in the case of the $\Lambda$CDM cosmology without primordial magnetic fields
and involves both the growing and decaying modes of
primordial fluctuations, which are proportional to $t^{2/3}$ and $t^{-1}$ 
in the matter dominated epoch, respectively.
Meanwhile, $D_{{\rm M} \rm p}(t)$ describe the growth rate of density fluctuations 
produced by primordial magnetic fields.
Assuming the matter dominated epoch, we can write $D_{{\rm M} \rm p}$ as
\begin{equation}  
D_{{\rm M}\rm b}(t) ={\Omega _{\rm b} \over \Omega _{\rm m}}
\left[
{9 \over 10} \left({t \over t _{\rm i} }\right)^{2/3}
 +9{\Omega _{\rm d} \over \Omega _{\rm b}} \left({t \over t _{\rm i} }\right)^{-1/3}
+{3 \over 5} \left({t \over t _{\rm i} }\right)^{-1} 
-{3 \over 2} \left({\Omega_{\rm m}+5\Omega _{\rm d} \over \Omega _{\rm b}}\right)
+3 {\Omega _{\rm d} \over \Omega _{\rm b}} \log \left({t \over t _{\rm i} }\right) \right],
\label{eq:mag-b-part}
\end{equation}
\begin{equation}  
D_{{\rm M}\rm d}(t) ={\Omega _{\rm b} \over \Omega _{\rm m}}
\left[
{9 \over 10} \left({t \over t _{\rm i} }\right)^{2/3}
-9 \left({t \over t _{\rm i} }\right)^{-1/3}
+{3 \over 5} \left({t \over t _{\rm i} }\right)^{-1} +{15 \over 2}
-3 \log \left({t \over t _{\rm i} }\right) \right].
\label{eq:mag-d-part}
\end{equation}
We plot the growth rates, $D_{{\rm M} \rm b}$ and $D_{{\rm M} \rm d}$, 
during the matter dominated epoch in Fig.~\ref{fig:transfer}.
We also show the growth rate of the density contrast for total matter,
$\delta_{\rm t}= (\rho_{\rm b} \delta_{\rm b} + \rho_{\rm d} \delta_{\rm d}) 
/(\rho_{\rm b} + \rho_{\rm d})$.  
In this figure, we normalize growth rates as $D_{{\rm M} \rm b}=1$ at $z=1$.
The figure shows that the density fluctuations of baryons are produced by the Lorentz force at first, 
while the density fluctuations of dark matter follow those of baryons gravitationally.
The growth rates of both fluctuations are proportional to $1+z$.

Next, we calculate the power spectrum of the density fluctuations.
Taking the assumption that there is no correlation 
between primordial magnetic fields 
and primordial density fluctuations for the sake of simplicity, 
we can describe the power spectrum as
\begin{equation}
P_{\rm p}(k) = P_{{\rm S} \rm p}(k)+P_{{\rm M} \rm p}(k) 
\equiv \langle |\delta_{{\rm S} \rm p}(k)|^2 \rangle + \langle |\delta_{{\rm
  M} \rm p}(k)|^2 \rangle, 
\label{matter-power}
\end{equation}
where $\delta_{{\rm S} \rm p}(k)$ and $\delta_{{\rm M} \rm p}(k)$ are Fourier
components of each density contrast.
The power spectrum $P_{{\rm M} \rm p}(k)$ is written as 
\begin{equation}
P_{{\rm M} \rm p}(k)     = 
\left ( t_{\rm i}^2 \over 4 \pi \rho_{{\rm b}0}a^3 (t_{\rm i}) \right)^2
D_{{\rm M} \rm p}(t)^2 I^2 (k),
\label{power-mag-part}
\end{equation}
where 
\begin{equation}
I^2 (k) \equiv \langle |\nabla \cdot (\nabla \times {\bm B}_0 ({\bm
x})) \times {\bm B}_0 ({\bm x})|^2 \rangle.
\label{nonline-convo}
\end{equation}

The isotropic Gaussian static of primordial magnetic fields
makes the nonlinear convolution Eq.~({\ref{nonline-convo}}) rewritten as 
\citep{wasserman, kim-olinto}
\begin{equation}
I^2 (k) = \int dk_1 \int d \mu {B^2 _{\rm n} (k_1) B^2 _{\rm n} (|{\bm k} -{\bm k}_1|)
\over |{\bm k} -{\bm k}_1 |^2 } (2 k^5 k_1^3 \mu+ k^4 k_1 ^4 (1-5 \mu^2) +
2 k^3 k_1^5 \mu^3),  
\label{nonline-convo-2}
\end{equation}
where $\mu$ is $\mu = {\bm k} \cdot {\bm k}_1/ |{\bm k}||{\bm k_1}|$.
Note that the range of integration of $k_1$ in Eq.~(\ref{nonline-convo-2}) depends on $ k$ 
because we assume that the power spectrum has a sharp cutoff below $1/k_{\rm c}$
so that $k_1 <k_{\rm c}$ and $|{\bm k} -{\bm k}_1 |<k_{\rm c}$ must be satisfied. 

We introduce an important scale for the evolution of
density perturbations, i.e., magnetic Jeans length.  
Below this scale, the magnetic pressure gradients, which we do not
take into account in Eq.~(\ref{eq:baryon-den}), 
counteract the gravitational force 
and prevent further evolution of density fluctuations. 
The magnetic Jeans scale is evaluated as \citep{kim-olinto} 
\begin{equation}
k_{\rm MJ}    
= \left[ 13.8 \left ( {B_{\lambda} \over 1 {\rm nG}} \right ) ^{-1}
\left ( { h^2 \Omega _{\rm m} \over 0.18} \right )^{1/2} 
\right ]^{2/n+5} {\rm Mpc}^{-1} .
\label{jeans}
\end{equation}
For simplicity,
we assume that the density fluctuations do not grow
below the scale,
although the density fluctuations below the scale are, in fact, oscillating 
like the baryon oscillation.

\begin{figure}
  \begin{center}
    \includegraphics[keepaspectratio=true,height=50mm]{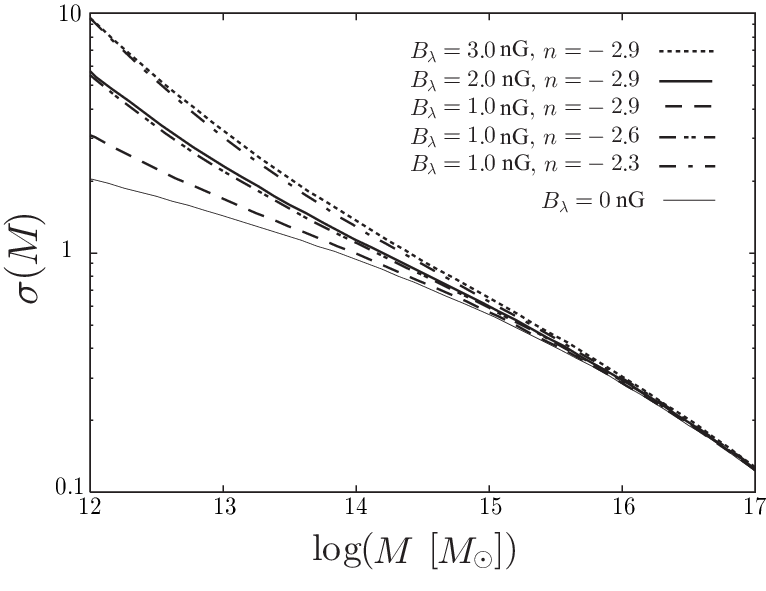}
  \end{center}
  \caption{Mass dispersion $\sigma$ for different primordial magnetic fields.
  The dotted, solid and dashed lines represent $\sigma$ for primordial magnetic fields
  with $B_\lambda=3.0$ nGauss, $B_\lambda=2.0$ nGauss, and
  $B_\lambda=1.0$ nGauss, respectively. Their power law indices are 
  $n=-2.9$. We also plot $\sigma$ for primordial magnetic fields with different power law indices;
  for $n=-2.6$ and $B_\lambda=1.0$ nGauss as the dashed-dotted-dotted line and 
  for $n=-2.3$ and $B_\lambda=1.0$ nGauss as the dashed-dotted line.
  For a comparison, we give $\sigma$ in the case without primordial magnetic fields as the thin
  solid line. }
  \label{fig:newsigma}
\end{figure}

In Fig.~\ref{fig:newsigma}, 
we show the mass dispersion $\sigma$, which is calculated from the power spectrum
of dark matter by
\begin{equation}
\sigma ^2(M) = \int dk k^2 P_{\rm d}(k) W(k R) ,
\label{eq:massdis}
\end{equation}
where $R$ is the scale which corresponds to mass $M$ 
and $W(x)$ is the top-hat window function. Here we normalized the primordial
matter fluctuations as $\sigma_8=0.8$.
The power law index of $\sigma$ does not depend on that of primordial magnetic fields.
This independence is brought by the sharp cutoff of magnetic fields
and the nonlinear term given by Eq.~(\ref{nonline-convo-2}).
We can analytically estimate Eq.~(\ref{nonline-convo-2}) in the limit
of $k/k_ {\rm c} \ll 1$ as 
$I^2(k) \sim \alpha B_{\rm c}^{2n+10} k^{2n+7} +
\beta B_{\rm c}^7 k^{4}$
where $\alpha$ and $\beta$ are coefficients which
depend on $n$~\citep{kim-olinto}.  Here we employ the
fact that the cutoff scale $k_{\rm c}$ is proportional to $B^{-1} _{\rm c}$
as is shown in Eq.~(\ref{eq:cutoff}). 
\citet{kim-olinto} found that the former term dominates if
$n < -1.5$, while the latter dominates for $n > -1.5$.  
However, we can find that 
the dispersion of the primordial magnetic fields 
with $n=-2.9$ and $B_\lambda=3.0$ and with $n=-2.3$ and $B_\lambda=1.0$,
or with $n=-2.9$ and $B_\lambda=2.0$ and with $n=-2.6$ and $B_\lambda=1.0$, are similar in Fig.~\ref{fig:newsigma}.
The magnetic fields of these pairs have almost the same Jeans scales.
Since the primordial magnetic fields have steep power spectrum, 
the contribution from the magnetic Jeans scale is large.
As a result, $\sigma_8$ of the magnetic fields with the same Jeans
scale (or cutoff scale) is almost same.

\begin{figure}
  \begin{center}
    \includegraphics[keepaspectratio=true,height=50mm]{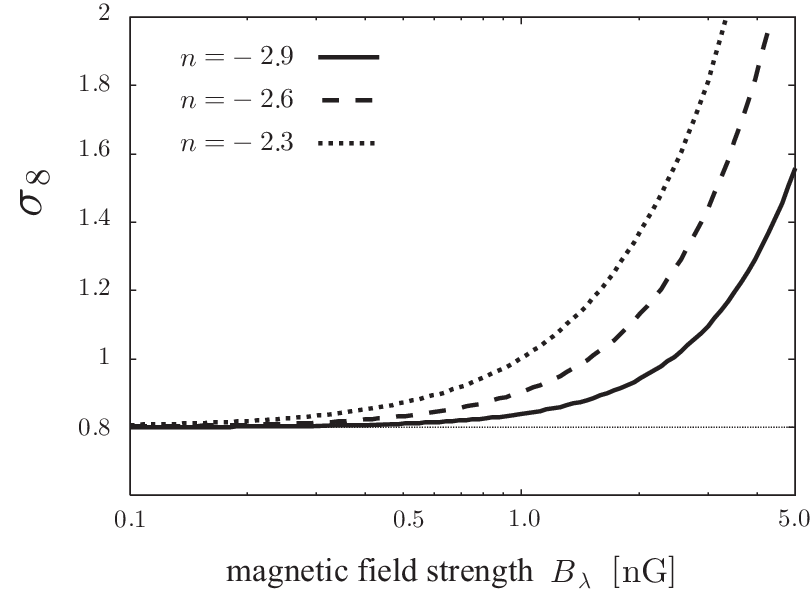}
  \end{center}
  \caption{Dependence of $\sigma_8$ on $B_\lambda$. 
  The solid line indicates
  $\sigma_8$ for primordial magnetic fields with $n=-2.9$.
  The dashed and the dotted lines represent $\sigma_8$ 
  for primordial magnetic fields 
  with $n=-2.6$ and with $n=-2.3$, respectively.}
  \label{fig:sigmamag}
\end{figure}

We also show $\sigma_8$ for different power law indices of primordial magnetic fields in 
Fig.~\ref{fig:sigmamag}.
In this figure, we plot $\sigma_8$ as the functions of $B_{\rm \lambda}$.   
The more blue spectrum primordial magnetic fields have,
the more amplitude of $\sigma_8$ they produce,
even if magnetic fields have the same strength at a given scale, 
for example, $h^{-1}$ Mpc in this paper.

In the calculation of the mass dispersion $\sigma$, we utilized
the top-hat window function. However, $\sigma$ depends on the choice
of the window function. The top-hat window function falls off as
$1/(kR)^2$ in large $k$. Because the power spectrum induced by the
primordial magnetic fields is very steep, some contribution for 
$\sigma$ comes from the magnetic Jeans scale. 
On the other hand, in the case
of the Gaussian window function, the contribution from the magnetic
Jeans scale is negligibly small due to the sharp cut-off of the 
window function on small scales. As a result, the amplification of 
$\sigma$ by primordial magnetic fields for the Gaussian window
function is much smaller than the one for the top-hat window 
function. Moreover, the critical density contrast $\delta_c$ and the
relation between window radius $R$ and mass $M$ depend on the choice
of the window function. \citet{lacey-cole-94} have compared the
analytic PS mass function with N-body simulations in the case of the
standard initial matter power spectrum, and found that, while
$\delta_c$ and the relation between $R$ and $M$ are independent on 
the spectral index of the power spectrum for the top-hat window
function case, they depends on the spectral index for the Gaussian
window function case. Therefore, N-body simulations with primordial
magnetic fields is necessary for detailed study about the effect 
of primordial magnetic fields. However it is beyond the scope of 
this paper.

\section{angular power spectrum of the S-Z effect}

The angular power spectrum of the S-Z effect is obtained through 
the halo formalism by many authors, e.g., 
\citet{cole-kaiser-88, makino-suto-93, komatsu-kitayama-99, komatsu-seljak-02}.
The angular power spectrum is given by
\begin{equation}
C_l  = g_\nu ^2 \int_0^{z_{\rm rec}} dz \frac{dV}{dz}
\int dM
{dn(M,z) \over dM} \left|y_l(M,z)\right|^2,
\label{eq:cl-sz}
\end{equation}
where $g_\nu$ is the spectral function of the S-Z effect which is $g_\nu =-2$
in the Rayleigh-Jeans limit, $V(z)$ is the comoving volume,
$n(M,z)$ is the comoving number
density of the dark matter halo with mass $M$ at redshift $z$, and $y_l(M,z)$ is 
the 2-D Fourier transform of the projected Compton $y$-parameter.
Presently, we are interested in multipoles higher than $l=300$,
and neglect the halo-halo correlation term in Eq.~(\ref{eq:cl-sz}).

For calculating ${dn(M,z) / dM}$ in Eq.~(\ref{eq:cl-sz}),
we adopt the Press-Schechter theory \citep{press-schechter-74},
\begin{equation}
{dn(M,z) \over dM} = \sqrt{2 \over \pi} {\bar \rho \over M} 
\left(-{\delta_c \over \sigma(M,z)} {\partial \sigma \over M} \right)
\exp \left(- { \delta_c ^2 \over 2 \sigma (M,z) }\right),
\label{eq:press-schechter}
\end{equation}
where $\delta_c$ is the critical over density.
The effect of primordial magnetic fields is taken into account 
through $\sigma (M,z)$ which is obtained from Eq.~(\ref{eq:massdis}) in the former section.

The 2-D Fourier transform component $y_l$ is given
in terms of the radial profile of the Compton $y$-parameter $y(x)$
through the Limber approximation,
\begin{equation}
y_l = {4 \pi r _{\rm s} \over l_{\rm s} ^2} 
\int ^\infty _0 dx x^2 y(x) {\sin(lx/l_{\rm s}) \over lx/l_{\rm s}}.
\end{equation}
Here $x$ is a non-dimensional radius $x \equiv r /r_{\rm s}$ 
where $r_{\rm s}$ is a scale radius which characterizes the radial profile,
and $l_{\rm s}$ is the multipole corresponding to $r_{\rm s}$.
The scale radius $r_{\rm s}$ is associated to the virial radius
with the concentration parameter $c$.
Following \citet{komatsu-seljak-02},
we set
\begin{equation}
c \approx \frac{10}{1+z}\left[\frac{M}{M_*(0)}\right]^{-0.2},
\label{eq:concentrait}
\end{equation}
where $M_*(0)$ is a solution to
$\sigma(M)=\delta_c$  at the redshift $z=0$.

As the radial profile $y(x)$,
we adopt the results of \citet{komatsu-seljak-02}.
They obtained $y(x)$ based on the NFW dark matter profile,
taking the three assumptions: the gas pressure and 
the dark matter potential reach the hydrostatic equilibrium; 
the gas density follows the dark matter density in the 
outer parts of dark halos; and the equation of state of gas
is polytropic $P_{\rm gas}\propto \rho_{\rm gas}^{\gamma}$
where $P_{\rm gas}$, $\rho_{\rm gas}$ and ${\gamma}$ are the gas pressure,
the gas density and the polytropic index.
According to these assumptions,
the radial profile $y(x)$ is written as
\begin{eqnarray}
y(x) &\equiv& {\sigma_T k_B \over m_e }n_e(x) T(x)
\nonumber \\
&=& {\sigma_T k_B \over m_e }n_e(0) T(0) y_{\rm gas} (x),
\label{eq:y-para}
\end{eqnarray}
where the gas profile $y_{\rm gas} (x)$,
the central number density $n_e(0)$ 
and the central temperature $T(0)$
are represented as
\begin{equation}
y_{\rm gas} (x)= 
\left\{
1 - 
 3\frac{\gamma-1}{\eta_{\rm c} \gamma}
\left[\frac{\ln(1+c)}c-\frac1{1+c}\right]^{-1}
\left[1-\frac{\ln(1+x)}x\right]
\right\}
^{1/\left(\gamma-1\right)},
\label{eq:gasprofile}
\end{equation}
\begin{equation}
n_e(0)
=3.01      
\left( {M \over 10^{14} M_\odot } \right)
\left( {r_{\rm vir} \over 1~ {\rm Mpc}  } \right)^{-3}
\left( {\Omega_{\rm b} \over \Omega_{\rm m}} \right)
\frac{c^2}{y_{\rm gas} (c) (1+c)^2}
\left[\ln(1+c)-\frac{c}{1+c}\right]^{-1}
~{\rm cm}^{-3},
\label{eq:rhog0}
\end{equation}
\begin{equation}
T(0) = 0.88 \eta_0
\left( {M \over 10^{14} M_\odot } \right)
\left( {r_{\rm vir} \over 1~ {\rm Mpc}  } \right)^{-1}
{\rm keV}.
\end{equation}
Here, the polytropic index $\gamma$ and 
the mass temperature normalization factor at the center $\eta_{\rm c}$ are given by
\begin{equation}
\gamma= 1.137 + 8.94\times 10^{-2}\ln(c/5) - 3.68\times 10^{-3}(c-5),
\label{eq:gammafitting}
\end{equation}
\begin{equation}
\eta_{\rm c}= 2.235 + 0.202(c-5) - 1.16\times 10^{-3}\left(c-5\right)^2.
\label{eq:etafitting}
\end{equation}

\section{results and discussion}

\begin{figure}
  \begin{center}
    \includegraphics[keepaspectratio=true,height=50mm]{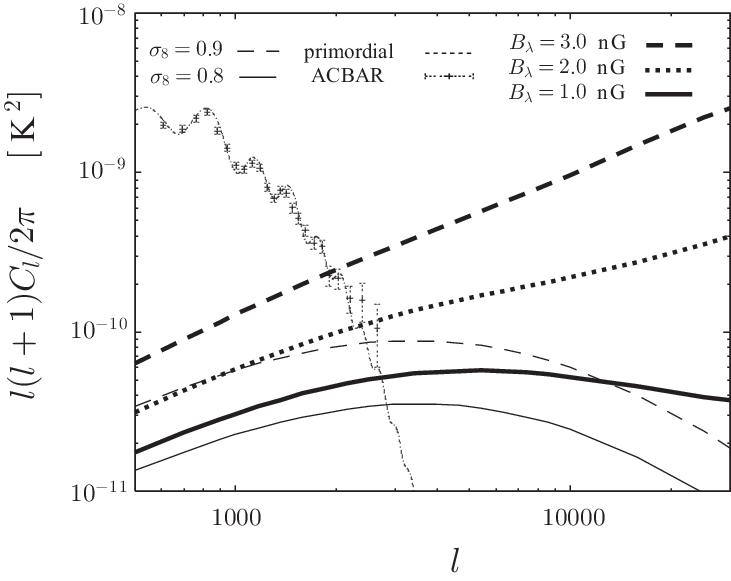}
  \end{center}
  \caption{ 
  S-Z angular power spectra for different magnetic field strength
  with $n=-2.9$. The solid line represents the S-Z spectrum 
  for primordial magnetic fields with $B_\lambda = 1.0$, 
  the dotted and the dashed lines indicate the spectra for $B_\lambda = 2.0$
  and $B_\lambda = 3.0$, respectively.
  The S-Z angular power spectrum without primordial magnetic fields 
  for $\sigma = 0.8$ and $\sigma = 0.9$ are shown
  as the thin solid line and the thin dashed line, respectively.
  For references, we plot primordial CMB temperature angular power spectrum and 
  ACBAR data.
  }
  \label{fig:clmag}
\end{figure}

First,
we calculate S-Z power spectra for different magnetic field strength with $n=-2.9$.
We plot the results on Fig.~\ref{fig:clmag}.
For references, we give the S-Z power spectra for the case of $\sigma_8=0.8$ and $\sigma_8=0.9$
without primordial magnetic fields.
We find the effect of primordial magnetic fields arises on small scales.
Although primordial magnetic fields with 2.0 $\mu$G  
amplify $\sigma_8$ to $0.9$ by the generation of additional density fluctuations
(see Fig.~\ref{fig:sigmamag}),
the S-Z power spectrum for 2.0 $\mu$G magnetic fields 
is much different
from that in the case of $\sigma_8 = 0.9$ without magnetic fields on small scales.
Therefore, the CMB observation on small scales has the potential to resolve
the degeneracy of $\sigma_8$ between the primordial density fluctuation and the additional 
density fluctuation by primordial magnetic fields.

\begin{figure}
  \begin{center}
    \includegraphics[keepaspectratio=true,height=50mm]{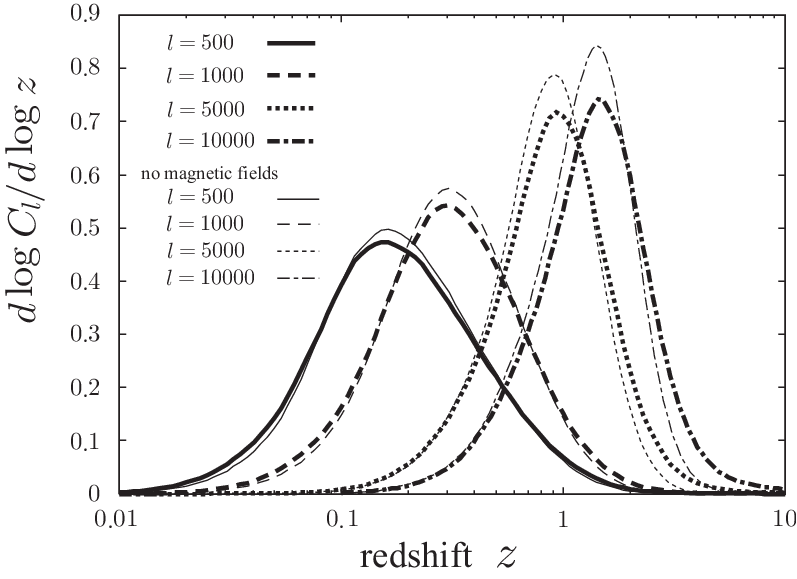}
  \end{center}
  \caption{
  Distribution of the redshift contribution of the S-Z angular power spectrum
  for given $l$ modes.
  Primordial magnetic fields have $n=-2.9$ and $B_\lambda = 1.0$. 
  The solid, the dashed, the dotted, and the dashed-dotted lines represent
  the distributions for $l=500$, $l=1000$, $l=5000$ and $l=10000$, respectively.
  For a comparison, we plot the distributions for the case 
  without primordial magnetic fields as thin lines.
  }
  \label{fig:redmag}
\end{figure}

The amplification of the S-Z power spectrum on small scales is due to
the early formation of dark halos which is induced by
the additional blue spectrum of the density fluctuations by primordial magnetic fields.
Since the electron density in Eq.~(\ref{eq:y-para})
is more dense in the early universe than in the late universe 
because of the cosmological expansion,
the S-Z power spectrum is more affected by high redshift structures
than other observations of mass distributions, for example, gravitational lensing.
Therefore, the early halo formation contributes to the amplification of the S-Z power spectrum
on small scales.
We can see this contribution in Fig.~\ref{fig:redmag}
where we show the redshift distribution of $C_l$ for given $l$ modes.
In large $l$ modes, 
there are enhancements in the tail part on the side of high redshifts
which come from the density fluctuations generated by primordial magnetic fields,
although the peak position is not changed, compared to the redshift contributions
in the case without primordial magnetic fields.

\begin{figure}
  \begin{center}
    \includegraphics[keepaspectratio=true,height=50mm]{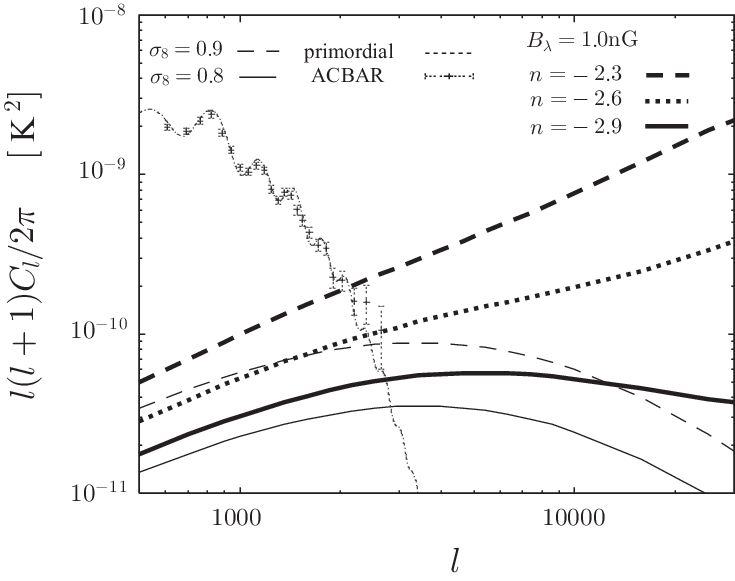}
  \end{center}
  \caption{
  S-Z angular power spectra for different power law indices.
  We choose $B_\lambda =1.0$ in all plots. 
  The solid, dotted, dashed-dotted, and dashed lines show the S-Z spectra for magnetic fields 
  with $n=-2.9$, $n=-2.6$ and $n=-2.3$, respectively.
  The S-Z angular power spectrum without primordial magnetic fields 
  for $\sigma = 0.8$ and $\sigma = 0.9$ are shown
  as the thin solid and thin dashed lines, respectively.
  For reference, we plot primordial CMB temperature angular power spectrum and 
  ACBAR data.
  }
   \label{fig:clnindex}
\end{figure}

Fig.~\ref{fig:clnindex} shows the S-Z angular power spectra for 
different power law indices of primordial magnetic fields. 
We choose $B_\lambda =1.0$ nGauss for all plotting cases. 
Comparing to Fig.~\ref{fig:clmag},
we find that the spectrum of primordial magnetic fields for $n=-2.6$ 
and $B_\lambda =1.0$ nGauss
is similar to that for $n=-2.9$ and $B_\lambda =2.0$ nGauss.
This is because, in the case of $n<-1.5$,
the dispersion of density perturbations caused by primordial magnetic fields depends more strongly on the cutoff scale of magnetic fields
than on the power law index, as mentioned in Sec.~\ref{density}.
Magnetic fields with $n=-2.6$ and $B_\lambda =1.0$ nGauss
and with $n=-2.9$ and $B_\lambda =2.0$ nGauss have almost the same cutoff scales 
so that they have similar S-Z angular power spectra,
even though their power law indices are different.

Although we computed the power spectra in the case of $n< -1.5$,
we will give some comment on the case of $n \ge -1.5$.
In such a case, the power spectral index of the density fluctuations generated by primordial magnetic fields
depends on $n$.
Therefore, the obtained S-Z power spectra with different $n$ are different,
even though the cutoff scales of primordial magnetic fields are the same.
The S-Z spectrum becomes steep if $n$ increases.

Since the S-Z power spectrum has a strong dependence on the cutoff scale
of primordial magnetic fields, 
we obtain the constraint on the cutoff scale 
by comparing with the observed CMB data on small scales.
For example, using the ACBAR data at $l=2500$ \citep{kuo-acbar}, 
we obtain $k_{\rm c} \la 95$ kpc.
This limit corresponds to $B_\lambda \la 2.0 $ nGauss
for $n=-2.9$ and $B_\lambda \la 1.0 $ nGauss for $n=-2.6$.
This result is comparable with other constraint given by other effects of primordial
magnetic fields on CMB temperature and polarization anisotropies caused by 
primordial magnetic fields, e.g., \citet{yamazaki-2008}.

\section{conclusion} 

We investigated the effect of primordial magnetic fields on the S-Z power spectrum.
Primordial magnetic fields generate additional density fluctuations after recombination,
so as to induce the early dark halo formation.
The generated dark halos in the early universe 
amplify the S-Z power spectrum on small scales.
We found that the amplification depends on the cutoff scale of primordial magnetic fields.
Therefore,
comparing our calculated results with present CMB observational data on small scales, 
we obtain the constraint on the cutoff scale of primordial magnetic fields,
$k_{\rm c} \la 95$ kpc.
This constraint is equivalent to
$B_\lambda \la 2.0 $ nGauss for $n=-2.9$
or $B_\lambda \la 1.0 $ nGauss for $n=-2.6$ at $h^{-1}$ Mpc.
The smaller the interesting scale of the S-Z power spectrum goes to,
the larger the enhancement by the primordial magnetic fields becomes.
Therfore we can expect that the future S-Z measurements can give  
constraints tighter than those from CMB temperature anisotropies
and polarization induced by the
magnetic fields at the recombination epoch.


The small scale CMB observations, e.g.,~CBI \citep{mason-cbi}, BIMA \citep{dawson-bima} 
and ACBAR \citep{kuo-acbar}
detected an excess of temperature anisotropies
from the small scale temperature anisotropy than what was expected from the WMAP results.
This excess corresponds to the S-Z effect with $\sigma_8=1.0$ \citep{bond-szsigma}.
However, this high value conflicts with the WMAP result, $\sigma_8 = 0.8$, 
which is obtained from the large scale temperature anisotropies \citep{wmap-spergel-07}.
The existence of primordial magnetic fields may resolve this discrepancy,
because the density fluctuations generated by primordial magnetic fields  
do not affect at large scales but add a blue spectrum on small scales.

The S-Z power spectrum depends on the electron density profile in dark halos.
For obtaining a highly accurate constraint on primordial magnetic fields,
we need a detailed study on the effect of primordial magnetic fields
on the electron density profile. 
However, we ignored this effect in this paper.
One possible effect of magnetic fields is brought by the pressure of magnetic fields.
The magnetic field pressure prevents electron gas from falling into 
the gravitational potential well of dark matter.
The modification of the electron density profile can be detected
by the S-Z effect,
if there are magnetic fields with several $\mu$Gauss in a halo  \citep{zhang-magsz}.
such magnetic field strength is easily obtained from primordial
magnetic fields with order of nano Gauss by adiabatic contraction
in the halo formation. 
We will study the effect on the S-Z effect due to primordial magnetic fields, 
and the consistent constraint on those fields,
considering effects other than 
the density fluctuations generation of primordial magnetic fields,
in the future.

\section*{Acknowledgements}
We would like to thank an anonymous referee for useful comments 
to improve our paper.
We also thank Richard Shaw for useful comments.
HT is supported by the Belgian Federal Office for Scientific,
Technical and Cultural Affairs through the Interuniversity Attraction Pole P6/11.
NS is supported by Grand-in-Aid for Scientific Research No.~22340056
and 18072004.
This research has also been supported in part by World
Premier International Research Center Initiative, MEXT,
Japan.


\end{document}